# Optical characterization of red blood cells from individuals with sickle cell trait and disease in Tanzania using quantitative phase imaging

JaeHwang Jung[1,*], Lucas E. Matemba[2,*], KyeoReh Lee[1], Paul E. Kazyoba[3], Jonghee Yoon[1], Julius J. Massaga[3], Kyoohyun Kim[1], Dong-Jin Kim[4,+], and YongKeun Park[1,5,+]

[1] Department of Physics, Korea Advanced Institute of Science and Technology, Daejeon 34141, Republic of Korea
[2] National Institute for Medical Research, P.O. Box 476, Morogoro, Tanzania
[3] National Institute for Medical Research, 3 Barack Obama Drive, P.O. Box 9653, 11101 Dar es Salaam, Tanzania
[4] Nelson Mandela African Institution of Science and Technology, School of Life Science and Bioengineering, P.O. Box 447 Arusha, Tanzania
[5] TomoCube, Inc., Daejeon 34051, Republic of Korea
*These authors contributed equally to this work.
Correspondence and requests for materials should be addressed to Y.K.P. (email: yk.park@kaist.ac.kr) or D.J.K. (email: djkim.nmaist@gmail.com)

**Sickle cell disease (SCD) is common across Sub-Saharan Africa. However, the investigation of SCD in this area has been significantly limited mainly due to the lack of research facilities and skilled personnel. Here, we present optical measurements of individual red blood cells (RBCs) from healthy individuals and individuals with SCD and sickle cell trait in Tanzania using the quantitative phase imaging technique. By employing a quantitative phase imaging unit (QPIU), an existing microscope in a clinic is transformed into a powerful quantitative phase microscope providing measurements on the morphological, biochemical, and biomechanical properties of individual cells. The present approach will open up new opportunities for cost-effective investigation and diagnosis of several diseases in low resource environments.**

Sickle cell disease (SCD) is an autosomal genetic blood disorder from the inheritance of point-mutated globin genes producing abnormal hemoglobin[1]. Under deoxygenated conditions, the abnormal hemoglobin, also known as hemoglobin S (HbS), become self-assembled inside red blood cells (RBCs), which results in the formation of rigid fibril structures. These fibril structures cause damages to the cell membrane making RBCs less deformable and can even change RBCs into sickle shapes. Stiffened RBCs in patients with SCD damage endothelial cells and even cause the occlusion of microvascular structures[2]. Thus, patients with SCD suffers severe anemia, pain, devastating disabilities, and in some cases, premature death[3,4]. In contrast, individuals with sickle cell trait (SCT), the heterozygous condition of SCD, do not exhibit apparent health issues. Without genetic analysis, the SCT individuals are hardly distinguished from normal individuals. Rarely, severe clinical manifestations including exertional rhabdomyolysis (the rapid breakdown of skeletal muscle due to injury to muscle tissue) have been reported in individuals with SCT under extreme conditions such as severe dehydration and high-intensity physical activity[5,6].

According to the World Health Organization, approximately 4.5% of the world population carries the sickle genes[7]. The sickle genes are found more frequently in the tropics, especially in Sub-Saharan Africa. For example, the prevalence of the sickle genes in Tanzania is estimated to be 13% and even up to 50% among some ethnic groups[8,9]. The mortality of infants with SCD is as high as 90% in areas with limited medical facilities and 50% in areas with improved health infrastructures[10] while only 1% of the infants with SCD dies in the United States[11]. The high prevalence of sickle genes imposes heavy economic and clinical burdens on Sub-Saharan Africa countries. Although genetic and biochemical information about SCD and SCT have been well understood, mechanical properties of these diseases have not been fully investigated. Measuring and understanding the mechanical properties of SCD and SCT are crucial to comprehend the mechanisms of diseases and evaluating the efficacy of drugs and medical treatments targeted to relieve the complications of the diseases. However, these kinds of studies have been mostly performed in developed countries in the US or Europe, mainly due to accessibility and well equipped medical research facizlities. Unfortunately, research in these developed countries may not reflect situations in Sub-Saharan Africa because it is difficult to find and access samples, and more importantly, the patients in these countries get reasonable medical care to relieve the symptoms and complications of SCD and SCT, such as treatment with hydrourexia.

Thus, the investigation of SCD and SCT in Sub-Saharan Africa can provide valuable information to understand the diseases. Despite the devastating burden of SCD, unfortunately, there have been insufficient investigations regarding SCT and SCD across the country due to the lack of funds, facilities, and experts. Hence, it is high time to develop and transfer simple, cost-effective and easy-to-use technology to help study the disease and build their knowledge about SCT and SCD. For a better understanding of SCD, SCT, and their complications, various technical approaches have been demonstrated mostly focusing on the mechanical properties of RBCs[12]. The mechanical properties of SCD RBCs have been measured based on invasive or force-applying techniques including micropipette aspiration[13] and filtration[14] as well as a flow-controlled chamber[15], optical tweezers[16], and atomic force microscope[17]. Although all these techniques have undoubtedly improved the understanding of the relation between SCD and the mechanical properties of the RBCs, simultaneous investigations of the multiple properties of individual cells are still required to better understand the diverse interactions of the mechanical, biochemical, and morphological characteristics.

Recently, quantitative phase imaging (QPI) techniques have been introduced to study the morphology and mechanical properties of individual SCD RBCs[18,19]. QPI is an interferometric microscopy technique capable of measuring optical phase delays of biological cells and tissues noninvasively and quantitatively[20,21]. However, most QPI techniques require complicated, sensitive, and bulky optical instrumentation which has to be operated and occasionally aligned by well-trained personnel. Therefore, all the previous studies with QPI have been performed at research facilities collaborating with medical hospitals in developed countries. This constraint is unfortunate because QPI has much to offer the fields of SCD and SCT with its unique single-cell profiling capability and label-free and quantitative imaging ability, particular for the study of RBC-related pathophysiology[22-28].

Here, we report the investigation of SCD and SCT RBCs in Africa using QPI. Exploiting a recently developed quantitative phase imaging unit (QPIU)[29], an existing simple bright field microscope in a local clinic in Tanzania was converted into a highly precise and sensitive QPI instrument. Using the QPIU, we performed optical measurements on the morphological, biochemical, and mechanical properties of individual RBCs collected from individuals with SCD and SCT. The quantitative characteristics of the RBCs including the aspect ratio, membrane curvature, hemoglobin (Hb) contents, and membrane fluctuation were measured from the quantitative phase images. In total, 585 RBCs from 23 individuals (5 healthy, 8 SCT, and 10 SCD) were measured, and their morphological, biochemical, and mechanical properties were systematically analyzed. We found that both the RBCs from SCD and SCT patients had significantly different morphological and biomechanical properties compared to the healthy RBCs. The demonstration of the QPIU in Tanzania will provide new opportunities for the cost-effective and quantitative investigation of SCD as well as other neglected tropical diseases in low- and middle-income countries.

**Results**

**Experimental setup**

To perform QPI in Tanzania, a standard bright-field optical microscope was converted into a highly stable and precise quantitative phase microscope with a QPIU (Fig. 1a). The QPIU is compact, simple, robust and easy-to-use. The QPIU consists of only three optical components: two linear polarizers and a Rochon prism (Fig. 1b). The unit can be attached to the image port of a standard microscope, and it does not require optical alignment. The principle behind the QPIU is lateral-shearing common-path interferometry[29]. Briefly, a beam from the image port of the microscope is first linearly polarized and then split into two beams with different propagation directions with a polarization prism (Fig. 1c). The polarization states of the two beams are matched to be parallel with the polarizer placed after the prism. Then, a spatially modulated interferogram is generated at an overlapped region of the two beams at the CCD plane (Fig. 1d). From the measured interferogram, an optical field image consisting of both the amplitude and phase map is retrieved using a field-retrieval algorithm[30].

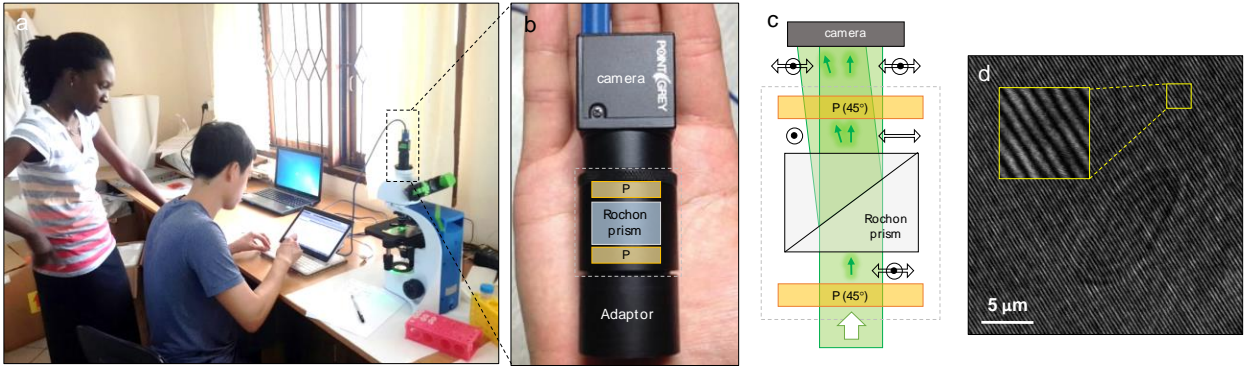

**Figure 1. Principle and demonstration of the quantitative phase imaging unit (QPIU).** (a) A photograph of experiments performed at a local clinic in Tanzania using the QPIU. (b) A photograph and (c) schematic of the QPIU. The QPIU consists of a Rochon prism and two parallel linear polarizers (P). The green and the black arrows indicate the optical paths and polarization direction, respectively. (d) A recorded hologram of a sickle-shaped RBC. The inset shows the magnified interference pattern.

In this experiment, a Rochon prism was used for the polarizing prism, instead of a Wollaston prism, which was used in a previous paper[29]. Whereas a Wollaston prism deviates both polarized beams from an optical axis, a Rochon prism deviates only one polarized beam. Thus, the use of a Rochon prism further simplifies the optical system of the QPIU because a CCD plane can be adjusted as perpendicular to the optical axis and not perpendicular to the direction of one of the deviated beams. The two polarizers are rotated so that the polarization direction of the polarizers have a 45° angle to the polarization axes of the beams, which ensures the maximum image contrast of the interferogram patterns and thus the maximum signal-to-noise ratio of the measurements. The Rochon prism and two polarizers are precisely adjusted and compactly assembled in an aluminum tube, which is mounted in front of a CCD camera (FL3-U3-32S2M-CS, Point Grey). As an illumination source, a laser diode (CPS532, $\lambda$ = 532 nm, Thorlabs Inc.) is installed on a bright-field microscope (B-382PLi-ALC, Optika) equipped with a 100× oil immersion objective (NA = 1.25, M-148, Optika).

The QPIU system is highly stable even when all the experiments were performed on a typical office desk. The temporal stability of the QPIU system, defined as the standard deviation value of background over time, was 9.4 nm, which is comparable to the 7 nm of fluctuation noise in a recent common-path interferometric system on a scientific-grade anti-vibration optical table[31]. This highly stable QPIU system is also cost-effective. The total cost of building a QPIU microscope is less than $3,000 USD including the microscope, the laser source, the camera and QPIU while a typical quantitative phase microscope system costs more than $30,000 USD. The cost only for the QPIU is less than $1,000 USD.

**Classification of the RBCs**
RBCs were collected at a medical center in Tanzania from 23 individuals (5 healthy, 8 SCD, and 10 SCT). At least 17 RBCs per participant and a total of 585 RBCs were measured to reduce cell-to-cell variations. The classifications of individuals into healthy, SCD, and SCT were confirmed with Genotyping by sequencing with polymerase chain reaction (See *Methods*). We categorized the collected RBCs into four groups: healthy RBCs, SCT RBCs, reversibly sickled SCD cells (RSCs), and irreversibly sickled SCD cells (ISCs). The healthy and SCT groups consisted of RBCs from healthy and SCT individuals, respectively.

The RBCs from individuals with SCD were separated into the RSC and ISC groups depending on their shapes. Because the experiments were performed under oxygenated conditions, sickle-shaped RBCs indicate that the cells have been permanently transformed (i.e., irreversibly sickled). The other RBCs from individuals with SCD were in discocyte shapes and considered as RSCs because the SCD patients in this study had not received a blood transfusion. The distinction of the ISCs and RSCs was achieved by visual inspections done by a trained medical doctor. Even though RBCs from SCD patients can be further classified into four subtypes based on their Hb concentration[32,33], we used the four classifications mentioned above mainly due to the lack of equipment for the separation in the clinic.

**Topography of the RBCs**

To demonstrate the capability of precise individual cell imaging, we performed quantitative phase imaging of the collected RBCs at the medical center in Tanzania. For each cell, 240 interferograms were recorded at a frame rate of 60 Hz with the QPIU. The optical phase image $\Delta\phi(x,y,t)$ is retrieved from a measured interferogram using a field retrieval algorithm[30]. Then, the hight map of a cell $h(x,y,t)$ is calculated from $\Delta\phi(x,y,t)$ using the relation $\Delta\phi(x,y,t) = 2\pi/\lambda \cdot \Delta n\, h(x,y,t)$, where $\lambda$ is the wavelength of light in a vacuum, and $\Delta n$ is the refractive index contrast between a RBC and the medium (See *Methods*).

The representative cell height maps are shown in Figs. 2a–d for the healthy, SCT, ISC, and RSC groups, respectively. Healthy RBCs exhibited characteristic discocyte shapes (Fig. 2a). SCT RBCs also showed discocyte shapes similar to the healthy RBCs (Fig. 2b). ISCs exhibited distinct sickle shapes whereas the RSCs had discocyte shapes (Figs. 2c–d).

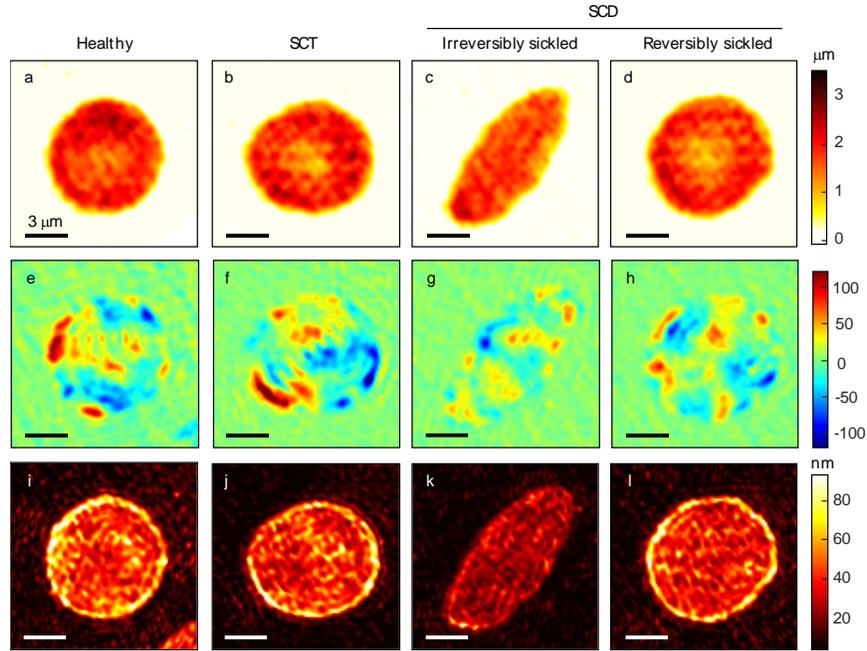

**Figure 2. Shapes and fluctuations of RBCs.** Height (the top row; a–d), instantaneous membrane displacement (the middle row; e–h), and standard deviation of displacements (the bottom row; i–l) map of a RBC from a healthy individual (a, e, i), a RBC from an individual with SCT (b, f, j), an irreversible sickle RBC (c, g, k) and a reversibly sickled RBC (d, h, l) from an individual with SCD, respectively.

**Dynamic membrane fluctuations of the RBCs**

To investigate the biomechanical properties of individual RBCs in SCD and SCT, we addressed the dynamic membrane fluctuations in the cell membranes. Dynamic membrane fluctuations in RBCs, with the displacement of tens of nanometers at millisecond temporal dynamics, are strongly related to the deformability of RBCs. These dynamic membrane fluctuations in RBCs have been quantitatively and precisely measured using QPI techniques[34-37], especially for the study of the effects of diverse pathophysiological conditions and the deformability of RBCs, including osmotic pressure[38], morphology[23], ATP[22], malaria[24,39,40], babesiosis[26], and cord blood[41]. Among them, QPI techniques have been previously used to measure dynamic membrane fluctuations in RBCs from patients with SCD[18,19]. We should note that the objectives of the this study are (1) to demonstrate the capability of QPI techniques in African countries regarding technical development and (2) to perform systematic measurements and comparative analyzes of individual RBCs from healthy, SCD, and SCT patients as scientific advances.

To address the dynamic membrane fluctuations in RBCs, dynamic displacements of the cell height maps $\Delta h(x,y,t)$ are obtained by subtracting the instantaneous height maps from the average height map as $\Delta h(x,y,t) = h(x,y,t) - \langle h(x,y,t)\rangle$. The representative instantaneous height displacement maps are shown in Figs. 2e–h for the healthy, SCT, ISC, and RSC groups, respectively. Healthy RBCs had high membrane fluctuations, implying large deformability (Fig. 2e), and SCT and RSC RBCs also showed comparable but slightly decreased deformability (Figs. 2f, h). However, ISCs exhibited significantly decreased dynamic membrane fluctuations, compared to the other groups, suggesting reduced deformability (Fig. 2g). Corresponding standard deviation maps of the dynamic membrane fluctuations were calculated and presented in Figs. 2i–l.

**Analysis of the morphological and biomechanical properties of the RBCs**

To systematically investigate the alterations associated with SCT and SCD, the morphological and biomechanical parameters of the RBCs were retrieved from individual RBCs, including the aspect ratio, cell membrane curvature, and membrane fluctuations (Fig. 3). The aspect ratio was calculated by the ratio of diameters along the long and short axes of a cell. The healthy and SCT RBCs had symmetric shapes. The ISCs presented significantly decreased aspect ratios due to the sickled shapes. The RSCs showed slight decreases in the aspect ratios compared to the healthy and SCT RBCs. The mean values of the aspect ratios were 0.89 ± 0.08, 0.90 ± 0.06, 0.44 ± 0.07 and 0.81 ± 0.12 (mean ± standard deviation) for the healthy, SCT, ISC, and RSC groups, respectively (Fig. 3a).

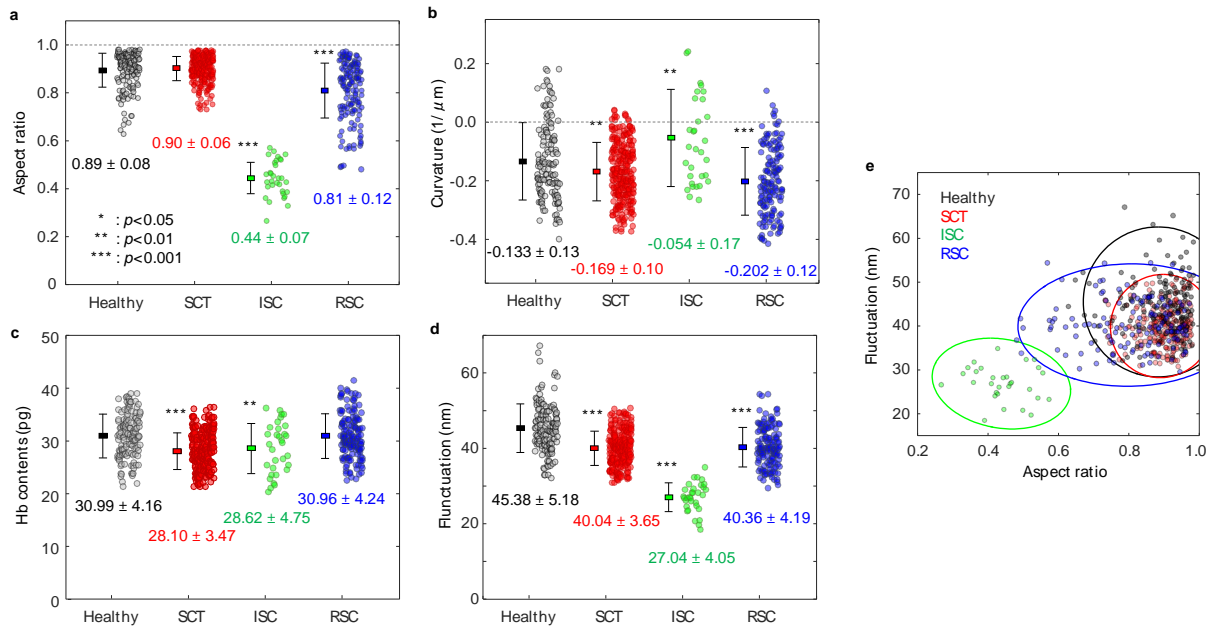

**Figure 3. Measured morphological and mechanical parameters.** (a) Aspect ratios of the projected areas of the cells. (b) Membrane curvatures of the center dimple regions. (c) Hb contents in individual cells. (d) Dynamic fluctuations in the cell membranes. (e) A scatter plot of the dynamic membrane fluctuation versus cell aspect ratio. Solid lines correspond to 95% confidence ellipses for each group. Statistical tests were performed with the Wilcoxon–Mann–Whitney tests (a) and Student's $t$-tests (b-d) against the Healthy group (*: $p<0.05$; **: $p<0.01$; ***: $p<0.001$).

To study whether a dimple shape is maintained, the curvature of the cell membrane was calculated from the measured cell height map (See *Methods*). The dimple shape in healthy RBCs was maintained with the intact spectrin cytoskeletal structures and membrane-bound proteins. It is known that polymerized HbS can cause damages to the cortex structures of the RBC membrane[42], the damaged membrane could result in the change of a dimple shape The mean values of the curvature are −0.133 ± 0.13, −0.169 ± 0.10, −0.054 ± 0.17 and −0.202 ± 0.12 μm$^{-1}$ for the healthy, SCT, ISC, and RSC groups, respectively (Fig. 3b). The results show that the SCT RBCs and RSCs have dimple shapes in the middle of the cell membrane, whereas the ISCs lost their dimple shapes.

**Hemoglobin contents in the individual RBCs**

The total mass of Hb inside RBCs is an important biochemical parameter indicating the capacity of oxygen transportation by the RBCs. Using the present method, the total amount of Hb inside individual RBCs, or the Hb contents, can be calculated from the quantitative phase images[43,44]. Because Hb represents the majority of the soluble proteins in RBCs, the Hb mass inside a cell can be related to the integration of the optical phase delay over the cell area as follows:

$$Hb \text{ contents} = C_{Hb} \int h(x,y) dA = \frac{\lambda}{2\pi\alpha} \int \Delta\phi(x,y) dA, \qquad (1)$$

where $\alpha$ is the refractive index increment of Hb (0.002 dL/g)[45]; $C_{Hb}$ is the Hb concentration in a RBC and $\int dA$ is the integration over the cell area $A$[46].

The mean values of the Hb contents were 30.99 ± 4.16, 28.10 ± 3.47, 28.62 ± 4.75 and 30.96 ± 4.24 pg for the healthy, SCT, ISC, and RSC groups, respectively (Fig. 3c). The mean Hb contents averaged over the RBCs correspond to the mean corpuscular hemoglobin (MCH), which is a medical parameter extensively used in the clinical laboratory. The mean value of the Hb contents of the healthy group was within the physiological level (27–33 pg). The Hb contents of the SCT and ISC groups were slightly decreased compared to the healthy RBCs, and the mean values of the Hb contents for every group were within the normal physiological range (27–33 pg). This result is in agreement with previous reports that there is no medically significant difference in Hb contents for SCT and SCD RBCs[47,48].

**Cell membrane deformability of individual RBCs**

To analyze the biomechanical properties of the RBCs, the dynamic membrane fluctuations of the RBCs in the four groups were quantified from the measured dynamic 2-D height fluctuation images. Dynamic fluctuation in RBC membranes, consisting of tens of nanometer displacements at millisecond temporal frequencies, reflect the biomechanical properties of RBCs, including the viscoelastic properties of the cell membrane cortex as well as the viscosity of the cytoplasm[49]. To represent the deformability of the RBCs, we calculated the temporal standard deviations of the dynamic membrane displacements and spatially averaged over a cell area as follows:

$$Fluctuation = \int \text{std}\{\Delta h(x,y;t)\} dA, \qquad (2)$$

where std{ } indicates the temporal standard deviation.

The average values for the fluctuations were 45.38 ± 5.18, 40.04 ± 3.65, 27.04 ± 4.05 and 40.36 ± 4.19 nm for the healthy, SCT, ISC, and RSC groups, respectively (Fig. 3d). The values of the membrane fluctuations for the healthy RBCs were the highest among the groups, and their values are comparable with previous reports[22-24]. The ISCs exhibited significantly decreased membrane fluctuations indicating a reduced cell deformability; the results agree well with previous reports[18,19]. The membrane fluctuations of the RSCs were between the fluctuation of the healthy RBCs and ISCs in agreement with previous findings[13,18,50]. The SCT RBCs had a similar level of fluctuations as the RSCs, which is also consistent with recent reports[16,51,52].

**Correlation between the shapes and deformability of the individual RBCs.**

The results above show that both elongated forms and reduced membrane deformability in RBCs are the distinct characteristics of the SCT and SCD. For further clarification, correlation analysis between the dynamic membrane fluctuation and the aspect ratio were performed at the individual cell level for the four groups (Fig. 3e). The ISCs population was clearly isolated from the other population, and the other groups shared overlaps in the scatter plot. Even though both SCT and RSC RBCs showed cell deformability comparable to the healthy RBCs, the SCT RBCs had aspect ratios compatible to the healthy RBCs whereas the RSC RBCs had slightly reduced aspect ratios. Additionally, in the scatter plot, the population of the SCT RBCs is a subset of the healthy RBCs, implying that the SCT RBCs are difficult to distinguish from healthy RBCs based on their morphological, biochemical, and mechanical properties.

**Discussion**

This study presents systematic cell analysis of RBCs in Tanzania using the QPIU. We performed optical measurements of multiple characteristics of RBCs collected from SCT and SCD individuals. To date, this is the first report to investigate using holographic microscopy or QPI techniques in a clinical setting in Africa. Despite the recent discoveries and developments of several biophotonic approaches targeted for translational medicine, the application to clinical studies in Africa has been difficult due to a lack of funds, facilities, and experts in African countries. This study performed in Tanzania clearly shows that QPI techniques can be transferred into local clinical environments in underdeveloped countries, which significantly expand the applicability and potentials of QPI techniques and provide new possibilities for research and diagnosis in environments with limited resources.

In this study, a systematic investigation of the morphological, biochemical, and mechanical properties of individual RBCs was performed. The collected RBCs were categorized into four groups, healthy, SCT, ISC, and RSC groups, depending on the genotypes of the donors and the cell morphology. For each RBC, the optical phase delay images were obtained, from which the morphological (cell aspect ratio and membrane curvature), biochemical (Hb contents), and biomechanical (dynamic membrane fluctuation) properties were retrieved and investigated.

The aspect ratios of the RBCs clearly showed the elongated shapes of the ISCs and the round shapes of the other RBCs. The RSCs had a broaden distribution for the aspect ratio which might be due to partially damaged membranes. The SCT RBCs showed almost identical round shapes with the healthy RBCs. In accordance with the similar shapes of the healthy and SCT RBCs and the RSCs, their membranes had the distinct curvatures of dimples. On the other hand, the membrane of the ISCs became flattened on average although some of the ISCs still exhibited strong dimpled membranes. The reduced curvatures of the ISCs were thought to be related to the damaged spectrin networks which are responsible for the dimple formation[53]. The detailed mechanism for the reduced curvature should be investigated with further experiments. For example, the QPI combined with optical tweezers could be used to study the relation between the membrane curvature and repeated elongation[54].

As a biochemical assessment, we measured the Hb contents inside the individual RBCs. The measured Hb contents of the SCT RBCs and ISCs were slightly lower (<10%) than that of the healthy RBCs. However, the deviations did not seem to be medically significant because all MCHs for the four groups were still within the physiological range. This result is consistent with previously work which had reported the MCH of SCT and SCD RBCs were within the physiological ranges[47,48]. Thus, the minor difference in MCH was believed to originate from the biological variations of the individuals. An additional source for the different MCHs could be related to the different losses of Hb components during a RBC life span[55]. An important fact is that the similar values of MCH among the groups clearly indicate the comparable oxygen capacity of the individual RBCs regardless of the sickle genes. Thus, it is evident that anemia is developed by the lack of an RBC population and blocked capillaries rather than the reduced oxygen capacity of the individual cells.

The membrane fluctuation, the dynamic characteristics indicating the deformability of the cells, was quantified by investigating the dynamics of the cell height. The measured membrane fluctuations showed that the SCT and SCD RBCs had less deformability than that of the healthy RBCs. In particular, the ISCs had the smallest fluctuation which was due to the permanent change in the membrane cortex structures as a result of repeated sickling[51,56]. It is worth noticing that the SCT RBCs and RSCs had lower values for the fluctuations compared to the healthy RBCs despite the indistinguishable morphology from the healthy RBCs. The result suggests that the reduced deformability not be solely caused by the polymerization of HbS and raises the necessity for further study on the mechanism of membrane stiffening.

From the correlative analysis between the aspect ratio and the fluctuation, we reconfirmed that the reduced deformability and elongated shape are distinct properties of the ISCs. The ISCs had the lowest fluctuation and aspect ratio among the groups, suggesting a correlation between the elongation and fluctuation. One possible reason is that a high aspect ratio means a high bending energy stored in the lipid membrane, which could suppress dynamic membrane fluctuation.

In this experiment, we categorized the SCD RBCs into two types, ISCs and RSCs, according to the shape of the cells. However, the RBCs from SCD individuals are often classified into four or more groups depending on their Hb concentration. With the combination of appropriate density-based separation techniques and QPI techniques, we would be able to investigate the characteristics of the RBCs according to more specific conditions. Furthermore, non-contact optical measurements of the individual membrane fluctuation controlling the oxygen level of the sample solutions may provide new quantitative findings for the hemodynamic characteristics of the SCT and SCD RBCs.

In summary, we have successfully demonstrated the practicality of the QPIU in Africa for the first time. We did measurements on the quantitative phase imaging of RBCs from individuals with SCD and SCT and investigated various cellular parameters. The method presented in this study can be readily applied in other clinical settings in underdeveloped countries because the QPIU is cost-effective and does not require precise optical alignment. This study also showed that an existing simple bright-field microscope could be converted into a high-performance quantitative phase imaging instrument capable of measuring morphological, biochemical, and mechanical properties at the same time. This approach provides valuable information to clinics which has been usually obtained with complete blood counter and cell biomechanics experimental apparatuses such as micropipette aspiration, which is difficult to implement and do in local clinics in underdeveloped countries.

We believe it is a perfect time to transfer the QPIU technique to underdeveloped countries as the awareness about the clinical and economic burdens of SCD increases. This demonstration suggests that the QPIU is an important and potential tool because the technique is cost-effective, compact, easy-to-use, and does not require expensive reagents or consumables. These advantages obviously provide additional opportunities for studying other parasitic and bacterial epidemics and will finally significantly contribute to the improvement of these health conditions in Africa. We believe that our demonstration in Tanzania presents an important and promising example of transferring advanced technology from the laboratory to the practical field.

**Materials and Methods**

**Recruitment procedures**
The study was conducted as a part of an ongoing hospital-based acute febrile illness study and done at the National Institute for Medical Research (NIMR) laboratory, located within Morogoro Regional Referral Hospital, Tanzania. Participants were recruited from the patients admitted to the pediatric ward number 6, and the remaining ones were patients who presented to the outpatient department with fever and other symptoms to seek treatment. Relatives of admitted participants also assisted in inviting other known sickle cell patients who were also requested to take part in the study. The remaining participants were healthy individuals who volunteered to take part in. Verbal consent was obtained from all study participants and the parents or guardian of recruited children before any sample collections. All study participants were tested for malaria using malaria rapid diagnostic tests (05FK60, Standard Diagnostics, Inc.) to minimize the chances of creating confusion with the effect of malaria. It has been suggested that malaria infection particularly *P. falciparum* parasites alters the characteristics of the cell membrane[24].

**Genotyping by sequencing polymerase chain reaction (PCR) products**
The sickle genotypes (AA, AS and SS) of the patients were confirmed by DNA sequencing of the PCR products of the beta-globin gene region including the sickle single nucleotide polymorphism (SNP)[57]. The genomic DNA was purified from 500 µl of whole blood with a commercial kit. The sickle SNP, A/T transversion in the codon 6 of the human beta-globin gene, was determined by DNA sequencing with the Big Dye Terminator sequencing kit (ABI). A 520 bp PCR product of the gene region including the sickle SNP loci was amplified with primers (Sickle_F: AAAGTCAGGGCAGAGCCATC; Sickle_R: AAGGGTCCCATAGACTCACCC), with 30 cycles of 94 °C for 30 sec., 55 °C for 30 sec., 72 °C for 60 sec. The corresponding PCR fragment was purified from the agarose gel and subjected to DNA sequencing.

**Sample preparations**
Under aseptic procedures, approximately 1 mL of peripheral blood was collected from the *participants* (known sickle cell patients and their biological parents and blood relatives). Specimens were collected in vacutainers tubes containing EDTA anticoagulant. All samples were stored at room temperature for a maximum of 10 hours before the measurements were conducted. The whole blood samples were diluted 500 times in phosphate buffered saline (PBS, Welgene Inc.) solution. The diluted blood solutions (5 µL) were injected into glass channels made of slide glass and a KOH-washed coverslip. Double-sided tapes of 100 µm thicknesses (3M Company) were used for spacers in the channels. The blood samples were considered to be under an oxygenated condition because no chemicals for deoxygenation were used, and the blood samples were sufficiently exposed to oxygen in the air.

**Calculating the RBC height from the optical phase images.**
The height of the RBC is obtained from $\Delta\phi(x,y,t)$ through the equation: $\Delta\phi(x,y,t) = 2\pi/\lambda \cdot \Delta n\, h(x,y,t)$, where $\lambda$ is the wavelength of light in a vacuum ($\lambda = 532$ nm) and $\Delta n$ is the refractive index contrast between an RBC and a medium (i.e., PBS solution). In this experiment, because the refractive index contrast can be considered to be solely originating from the Hb inside the RBCs, the refractive index contrast could be expressed as $\Delta n = \alpha \cdot C_{Hb}$[44,58], where $C_{Hb}$ is the Hb concentration, and $\alpha$ is the refractive index increment of the Hb ($\alpha = 0.002$ dL/g[45]). Because we were not able to measure $C_{Hb}$ for the individual RBCs at the clinic, the values of $C_{Hb}$ were adapted from previous reports to determine $\Delta n$. Specifically, for the healthy and SCT RBCs which have a physiological concentration of Hb, we used $C_{Hb} = 34$ g/dL[45,47]. For the SCD RBCs, two different $C_{Hb}$ values were used depending on the groups because the Hb concentration of the SCD RBCs is a bimodal distribution[32,59]. One population, mainly comprised of the RSCs, is known to have an average Hb concentration similar to that of the healthy RBCs ($C_{Hb} = 34$ g/dL).

The other population, mainly comprised of the ISCs, is known to have an average concentration of 45 g/dL[32]. Thus, the optical phase delay of 1 radian corresponds to the height of 1.249 μm for the healthy and SCT RBCs and RSCs and a height of 0.944 μm for the ISCs.

**Calculating the membrane curvature**

To quantify the membrane curvature of the individual cells, a mean curvature, which is the mean of two principal curvatures, at each point on the cells was calculated from the height map. The calculated mean curvatures at each point were averaged over a dimple region around the center of the cells. The dimple region is defined by an elliptical area whose diameters are 40% of the diameters of the individual RBCs.

**Acknowledgments**

This work was supported by KAIST, and the National Research Foundation of Korea (2015R1A3A2066550, 2014K1A3A1A09063027, 2012-M3C1A1-048860, 2014M3C1A3052537) and Innopolis foundation (A2015DD126). We thank all study participants who accepted to take part in this study.

**Author contributions**

Y.P and D.K conceived the idea and directed the work. J.J., L.E.M performed experiments and wrote the main manuscript. J.J, K.L, and J.Y. designed optics system and discuss results. P.E.K and J.J.M prepared RBC samples. All authors reviewed the manuscript.

**Competing Financial Interests statement**

The authors declare no competing financial interests.